\documentstyle{article}

\title{Self-Consistent Microscopic Model  of Fluctuation-Induced
Transport}   \author{Mark M.
Millonas  \\ \small Theoretical Division and CNLS,
MS B258 Los Alamos National Laboratory, \\ \small    Los Alamos, NM 87545
  }

\begin{document}

 \maketitle

\begin{abstract}
A Maxwell's demon type ``information engine" which extracts
work from a bath is constructed from a microscopic
Hamiltonian for the whole system including a subsystem, a thermal  bath, and a
nonequilibrium bath of phonons or photons which represents an
information source/sink. The kinetics of the engine is calculated
self-consistently from the state of the nonequilibrium bath, and the relation
of
this kinetics to the underlying microscopic thermodynamics is established.
\end{adstract}

Processes in which some of the energy in a nonequilibrium bath is transformed
into work at the expense of increased entropy are of great interest in
a number of important areas, but  the study of the
kinetics of such processes  is
complicated by the fact that no  principles of the power and generality of
those
of equilibrium statistical mechanics exist for such cases.
A number of semi-heuristic type models  have appeared recently which have
served
as illustrations that time correlated fluctuations interacting with a spatial
asymmetry are sufficient condition to give rise to
transport.\cite{groups,Me,ddd}  An application  of this idea has recently
been utilized experimentally as a new type of molecular separation
technique.\cite{exp} In addition, it is  clear that spatial asymmetry is  a
necessary requirement only when all the odd moment of the fluctuations
(including
orders higher than first) vanish, and that transport will generally occur even
in
the absence of a spatial asymmetry if this requirement is not met.\cite{cm}

These models,
which come under the general heading of ``fluctuation induced transport",  are
usually based on a reduced description of the noisy overdamped motion of a
particle in a periodic potential.  The nonequilibrium
effects of the irreversible ($\delta S >0$)  interaction of the system with a
nonequilibrium bath are modeled in various ways, and net current appears as a
consequence, even though the
average driving force vanishes. In this way these nonequilibrium fluctuations
can be used to do work.

This previous work has focused on phenomenology, and no attempt was made to
formulate self-consistent models.  Since the choice of the
kinetics of the reduced system is somewhat arbitrary, it is often difficult to
know whether such descriptions are appropriate, or even
allowed by the microscopic laws of physics. In order to treat  these models
 self-consistently reduced descriptions need to be  carefully
derived from microscopic considerations since
equilibrium kinetics is no longer applicable.
Here we construct a  special microscopic model which contains an
explicit description of the bath as well as the sub-system, and allows a
rigorous
determination of the kinetics. This model can be used to more
fully explore the question of what types of kinetic description are allowed by
the underlying laws of physics, and how these kinetic descriptions are related
to the state of the bath, and its' fundamental thermodynamic
properties.

We will consider a particle (sub-system) with position $Q$ coupled to a thermal
``Brownian" bath ${\cal A}$, which represents the thermal background
environment
of the engine,  and to a nonequilibrium bath ${\cal B}$.  As we will
demonstrate, {\it the non-thermal part of the energy in bath ${\cal B}$ can be
viewed as a source or sink of negentropy (physical information)  which allows
the
engine to operate, while the thermal parts of both baths provide the actual
energy, as in the case of Maxwell's demon.}  The Hamiltonian for the entire
system will be given by   \[{\cal H} = {M\over 2}\dot Q^2  +U(Q) +{\cal
H}_{{\cal
A}}+{1\over 2} \sum_k (\dot Y_k^2 + \omega_k^2 Y_k^2)  \]
\begin{equation}+{\cal
H}_{int\  {\cal A}}-\epsilon\  V(Q)\sum_{k}   Y_k. \end{equation} The first two
terms on the rhs  describe the subsystem, where $M$ is the mass of the
subsystem.
${\cal H}_{\cal A}$ is the Hamiltonian for the Brownian bath.  The fourth term
describes the  bath ${\cal B}$, which is represented as bath of linear
oscillators, with   frequency spectrum $\{\omega_k\}$. The last two terms are
the interaction of the subsystem with the baths, where $\epsilon$ is a coupling
constant.  The form of the nonequilibrium bath ${\cal B}$, that of a set of
phonons, was chosen for both simplicity, and because of its generic
relationship
to many condensed matter type systems.  Extensions of this approach to higher
dimension (more gross variables) are straightforward.

The evolution of  ${\cal B}$ is given by
\[
Y_k(t) =  A_k \cos(\omega_k t + \phi_k) \]
\begin{equation}
   + {\epsilon\over
\omega_k}\int_0^t d\tau\  V(Q(\tau))\sin
\omega_k(t-\tau),\end{equation}
where $A_k$ and $\phi_k$ are the initial amplitudes and phases of the
 oscillators.  This equation can be used to eliminate the oscillator modes, and
to obtain a description of the variable $Q$.\cite{z}
 We will assume that the
 interaction of the subsystem with ${\cal A}$ is that of a Brownian
particle, and that the frequency spectrum  of the oscillator bath ${\cal B}$
is  quasi-continuous with a
 frequency density
$\rho(\omega)$
of the Debye type
\begin{equation}
\rho(\omega) = \left\{ \begin{array}{lll}
{3\omega^2/ 2\omega_c^3} & \ \ \   & \vert\omega\vert\leq \omega_c \\
 0 & \ \ \   &\vert \omega\vert > \omega_c
\end{array}\right.
  \end{equation}
which is regularized by a cutoff at high frequency $\omega_c$ which
 is assumed to be larger than
any typical frequency of the gross variable.  Since the bath is quasi-infinite
we can assume that the state of the bath does not change on time
scales of interest as a result of its interaction with the subsystem. After
elimination of  the bath variables from the equations of motion we obtain a
nonlinear Langevin equation for the subsystem, \begin{equation}  M\ddot Q+
\Gamma(Q) \dot Q
+\tilde U^\prime(Q)  = \xi_{\cal A}(t) + V^\prime(Q) \xi_{\cal
B}(t) \end{equation} where $\Gamma(Q) = \Gamma_{\cal A} +[V^\prime(Q)
]^2\Gamma_{\cal B}$,  where $\xi_{\cal A}(t)$ is Gaussian white noise,
  \begin{equation}<\xi_{\cal A}(t)>=0,\ \ <\xi_{\cal A}(t)\xi_{\cal A}(s)> = 2
\Gamma_{\cal A} k T \delta(t-s), \end{equation}
and $\xi_{\cal B}(t)$ is a Gaussian noise with
\[
<\xi_{\cal B}(t)>=0,\ \ \ \phi(t) = <\xi_{\cal B}(t)\xi_{\cal B}(0)>.
\]
\begin{equation}
\Phi(\omega) = \int^\infty_{-\infty} dt\ \exp(i\omega t)\phi(t)
=4\Gamma_{\cal B}u(\omega)
\end{equation}
where $u(\omega) = < \omega^2 A^2(\omega)>/2$ is the energy density,
 which depend explicitly on the preparation of the bath.
In addition the ``bare" potential is now dressed by the oscillator bath, $
 \tilde U(Q) = U(Q)- (\omega_c/\pi) \Gamma_{\cal B} V^2(Q)
 $.  Here  for simplicity we assume a random distribution of initial phases
of the oscillators,  which insures that the noise is Gaussian.  Here the
only approximation that has been made in going from  Eqs. 1-3 to  Eqs. 4 and 5
is neglect of the Poincar\'e  recurrence time of the system, and Eq. 7 follows
from the random phase assumption.

 For the purposes of this paper we will consider only the
over damped ($\Gamma_{\cal A}/M >>1$) case, so
that \begin{equation}
\Gamma(Q) \dot Q =-
{\tilde U^\prime(Q) }+ \xi_{\cal A}(t) +
 {V^\prime(Q)} \xi_{\cal B}(t). \end{equation}
The inclusion of the thermal ``Brownian" bath ${\cal A}$ plays an important
role
here since this description will break down when $\Gamma_{\cal A}=0$. We will
use this equation to study  fluctuation induced transport in  a system where
$U
(Q) = U(Q+\lambda)$, and
 $V (Q) = V(Q+\lambda)$,
so that   the Hamiltonian is
invariant under the translation $Q\to Q+\lambda$.
As a consequence $\tilde U(Q) = \tilde
U(Q+\lambda)$. A portion of a typical ``dressed" ratchet potential $\tilde
U(Q)$
is pictured in Fig. 1.
 Even though the average force on the particle vanishes, a net current will be
produced, which if directed against a load
force can be used  to do work.     The basic theoretical problem is to find the
mean velocity $\langle \dot{Q}(t)\rangle$ in the sub-system given
the shape of $U(Q)$ and $V(Q)$, and the properties of the noise  terms
$\xi_{\cal A}(t)$ and $\xi_{\cal B}(t)$.

\begin{figure}[t]
\vspace{1.45in}
\hspace{0in}\special{postscriptfile pot scaled 900}
\caption{\sf Typical dressed ratchet potential $\tilde U(Q)$.}
\end{figure}

Since we have started with an explicit microscopic (time reversible)
Hamiltonian,
if the system as a whole is in equilibrium the current must vanish. Therefore a
stationary current
 can  appear  only if the system is out of equilibrium.  This is a
basic consequence of the second law of thermodynamics  which requires that no
net
work can be extracted from a system in thermal equilibrium.  Work can be
extracted
from the system via a  Carnot type engine  which runs off of
two baths at different temperatures.   Our system  can operate as such an
engine if ${\cal B}$  is prepared in a quasi-thermal state, that is, where
the temperature of ${\cal B}$ is not necessarily equal to the temperature
of the bath ${\cal A}$  ($T\neq \bar T$). The
then equipartition of energy then gives $u(\omega) =  k \bar T/2$,
$\xi_{\cal B}(t)$ is  Gaussian white noise with
$
<\xi_{\cal B}(t)>=0,$ and $\phi(t) = 2\Gamma_{\cal B} k \bar T\delta(t)$,
and  Eq. 7 is Markovian, and  thus amenable to
standard techniques.
 The evolution of the probability density
$\rho(Q,t)$ for the system described by Eq. 7 is then given by the
Fokker-Planck equation,
 \[
\partial_t \rho = \partial_Q\left[
 {\tilde U^{\prime}(Q)/\Gamma(Q)}
+kT{\partial_Q}({{\cal D}(Q)/ \Gamma(Q)})\right]\rho,
\]
\begin{equation}
{\cal D}(Q) = 1 +{r(\Gamma_{\cal B}/\Gamma_{\cal A})(V^\prime(Q))^2\over
1+ (\Gamma_{\cal B}/\Gamma_{\cal A})(V^\prime(Q))^2}, \end{equation}
where $r ={ (\bar T-T)/T}$.

  Eq. 8 can be solved for the
steady-state solution with periodic boundary conditions  $\rho_s(x) =
\rho_s(x+\lambda)$, and normalization $\int_x^{x+\lambda} \rho_s(x)\ dx
=1$.\cite{VK}  This yields an exact expression for the average velocity \[
<\dot Q> = {kT[1-\exp(\delta/kT)]\over
\int_0^\lambda dy\  e^{-\Psi(y)/kT}\int_y^{y+\lambda} dx\
\Gamma^2(x) e^{\Psi(x)/kT}/{\cal D}(x)} \]
\begin{equation}
\Psi(x) = \int^x {U^\prime(y)\over {\cal D}(y)} \ dy,\ \ \
 \delta = \Psi(0)-\Psi(\lambda).
\end{equation}
It is easy to see from Eq. 9  that when the temperature difference
between the baths is zero ($r = 0$), the current  vanishes identically
(since $\delta = 0$).  This is to be expected, and of course is
a consequence of the second law.  The current will also vanish in the limit
$\Gamma_{\cal B}/\Gamma_{\cal A}\to 0$.

 From this point on we will only  consider
 cases where
the characteristic noise intensities $T$ and  $D = {\rm max}\,\Phi
(\omega)$ are small in comparison to the well depth $\Delta \tilde U =
\tilde U(b)-\tilde U(a)$, which can be insured by making the coupling between
the
system and the bath small enough. This situation is particularly interesting
since analytic results are possible for both Markovian, and non-Markovian
situations,\cite{Me,mark} and since the basics physics is illustrated most
clearly.

For  $T,D<<\Delta \tilde U$   most of the time the system
performs small-amplitude fluctuations about the minima of the  potential.
Occasionally it will ``jump" from the minimum it  occupied to the  one on the
right or left, with the  probabilities per unit time $W_+$ and $W_-$,
respectively.  These jumps give rise to the average velocity $<\dot Q>=\lambda
(W_+-W_-)$.

For the Markovian case described in Eq. 7 the transition rates
can be calculated via standard techniques, and evaluated by steepest
descents.  We obtain $ W_\pm = W_K \exp(r \beta_\pm/kT)
 $
where
\begin{equation}
 W_K = { \sqrt{ U^{\prime\prime}(a)\vert U^{\prime\prime}(b) \vert}\over 2
\pi} \exp(-\Delta U/kT)
 \end{equation}
is the Kramers activation rate, with $\Delta U = U(b)-U(a)$, and where for
small $\Gamma_{\cal B}/\Gamma_{\cal A}$,
\begin{equation} \beta_\pm = (\Gamma_{\cal B}/\Gamma_{\cal A})\int_a^{b^\pm}
U^\prime(x) [V^\prime(x)]^2 \ dx .\end{equation}
These transition rates can be
further expanded in powers of $(\Gamma_{\cal B}/\Gamma_{\cal A})$, but for our
present purpose, this is not particularly enlightening. The mean velocity is
given by \begin{equation}  <\dot Q> =\lambda W_K\left[ e^{r \beta_+/ kT}-e^{r
\beta_- / kT}  \right].
 \end{equation}
This expression can also be obtained from the exact solution 9 by evaluating
the
integrals in the denominator via steepest descent.   We see that the current
will
flow in one direction if $T <\bar T$, and the opposite direction if $T >\bar
T$.  Thus, the system acts like as Carnot engine, doing work by making use of
two thermal baths at different temperatures.

The correlation ratchet, a system which is driven by the effects of colored
noise,
is obtained from Eq. 6 and 7 by setting $u(0) =kT/2$.  Thus both
${\cal A}$ and ${\cal B}$  have ``thermal parts" while ${\cal B}$ has a small
part $u(\omega)-u(0)$ which deviates from equilibrium.  For this case
the lowest order terms which produced the current will be independent of
$V(Q)$,
and depend only on the properties of $\xi_{\cal B}(t)$. If bath ${\cal B}$ has
a
non-thermal distribution over its modes, then $u(\omega)$ is not constant, and
this manifest itself as time correlations  (i.e.  $\xi_{\cal B}(t)$ is no
longer
delta correlated)  and     a net
current will arise.

When the bandwidth of the spectrum $\Phi (\omega)$ greatly
exceeds the reciprocal relaxation time of the system $t_r^{-1}
= \tilde U^{\prime\prime}(a)$,   the transition
probabilities $W_{\pm}$ can be calculated by an extension of the variational
technique used in \cite{mark,Me}, where
$
W_{\pm} = W_K \exp(-\gamma_\pm F^{\prime\prime}(0)/kT)
$ and
\begin{equation}
\gamma_{\pm}=\int_{a}^{b_{\pm}}
\tilde U^\prime(x)[\tilde U^{\prime\prime}(x)]^2 dx \geq 0,
\end{equation}and where
$F(\omega) = kT/4u(\omega)$,
$F^{\prime\prime}(\omega) = {d^2F(\omega)/ d\omega^2}$, with $
\vert F^{\prime\prime}(0)/F(0)\vert<<t^2_r$.
The mean velocity
\begin{equation}
<\dot Q> = \lambda W_K
 \left[e^{-
{\gamma_+F^{\prime\prime}(0)/
 kT}} -e^{-{\gamma_-
F^{\prime\prime}(0)/kT}}\right].
\end{equation}
 We have neglected the
small corrections to the prefactor in $W_{K}$ due to the noise
color and used the standard
Kramers expression for this prefactor valid for white-noise driven systems.
 The direction of the current is
determined by the interplay of the shapes of the potential and energy density
distribution $u(\omega)$. Just as the current in the thermal ratchet changes
sign
when $r$ changes sign, the current in the correlation ratchet changes sign when
$F^{\prime\prime}(0)$ changes sign.  More details can be found in \cite{Me}.
These current reversals are due to an activation effect, and are entirely
unrelated to the current reversals found in \cite{ddd}.

 Although the corrections
$\gamma_{\pm}F^{\prime\prime}(0)$ to the activation energy are small compared
to
the main term, they are not small compared to $kT$,
and can change $W_{\pm}$ by orders of magnitude.
Excepting the special case where
$\tilde U(Q)$ is symmetric with respect to $a$, the transitions in one
direction
will typically dominate overwhelmingly over the transitions in the opposite
direction. The optimal rate $<\dot Q>=\lambda
W_K$ is attained when all the thermally activated transitions are in one
direction. Thus, while the vast majority of the energy in  both ${\cal A}$
and ${\cal B}$ is  thermal distributed in this near-equilibrium
situation it is the relatively small amount of energy which is not distributed
thermally, or equivalently the negentropy, which allows the engine
to run.  On the other hand if the thermal energy  where removed the engine
would
immediately stop running since  virtually no transitions would ever occur. It
should be clear from previous analysis that at  the force driving the particle
comes overwhelmingly from the thermal parts of the baths.
Therefore we must conclude that
while even a very small negentropic source/sink in ${\cal B}$ allows the engine
to
operate, the thermal fluctuations provide the energy.

As described in the preceding paragraph this  system is   an
``information engine"  analogous to a Maxwell's Demon engine,  which extracts
work out of a thermal bath by rectifying the thermal fluctuations of the
system.
Maxwell's  Demon is a ``being"  which uses information about the system to
``choose" only those fluctuations which are helpful to make the engine run.
This
information, which can only be acquired if the Demon is not in equilibrium with
the bath,\cite{B} is   used to rectify the energy already available, but
otherwise
inaccessible, in the thermal bath.  As shown by Szilard\cite{Szilard}, the
information is acquired at the expense of an entropy increase of the Demon, an
observation which salvages the second law.  Similarly it is clear from the
approach used here that our system does work  at the expense of the total
increase of entropy of the baths, and operates because of the physical
information contained in the non-thermal energy of the bath, while the energy
 is paid predominately in the currency of the thermal fluctuations.

In the example given by Brillouin in \cite{B} the demon uses light photons to
determine the location of a particle, and then uses this information to extract
work from the system.  The demon needs a source of light which is not in
equilibrium with the bath in order to distinguish the signal from the thermal
background radiation.  The model presented here can be regarded as a
simplified picture of a bath of photons coupled to a particle in a thermal
bath.
By adding or removing photons (energy) from a system in thermal equilibrium an
information source/sink is created of the same type as described by  Brillouin.
The sub-system in this case plays  the role of the demon which allows the
information to be converted to work.

This observation is made precise in the following way.  Once the energy density
over the frequency spectrum of the phonon bath
$u(\omega)=<\omega^2 A^2(\omega)>/2$ is known,  thermodynamic quantities can
be calculated. Near equilibrium, as is the case for the above approximation,
nearly all of the energy in the two baths is in a thermal state, and any
entropy
increase $\delta S$ will not change the temperature.  In this case the
physical information (negentropy) in the phonon bath   is given isothermally
by    $H_b = \int_0^{\omega_c}d\omega\  u_i(\omega) $ where  $u_i(\omega) =
u(\omega)/T-k/2$ is the information density.
Since we have
set   $\Phi(0) =2\Gamma_{\cal B}kT$, the sign of the information contained in
the low frequency part of the spectrum is determined by the curvature of the
information density at zero frequency,  $u_i^{\prime\prime}(0)= - k
F^{\prime\prime}(0)$  as illustrated in Fig. 2. The situation
$F^{\prime\prime}(0)<0$ implies a low frequency ``source" of information in
${\cal B}$, and while $F^{\prime\prime}(0)>0$ a ``sink" in ${\cal B}$ as is
illustrated in Fig. 2. As we have already shown in the above, the engine will
run in opposite directions in these two cases. When $H_b >0$  information
flows out of ${\cal B}$ and the engine turns in one direction. The first is
just
thermodynamics, while the second is a result of the previous calculations. Just
the opposite is the case when $H_b<0$, and when the system is in equilibrium
$H_b =0$. Thus the semi-huristic treatments of \cite{mark,Me} can be made
self-consistent, and the relationship between thermodynamic quantities and
reduced kinetic descriptions such as Eq. 4  can be established.

The free energy of the system is given by ${\cal F} =\tilde U +T
H_b-TS_a$ where $S_a$ is the entropy of the thermal bath,  but in the
nonequilibrium case ${\cal F}$  is generally not sufficient to  calculate
rates,
as should be  clear from the above example.  While (near equilibrium) the free
energy does play the role of a stochastic Lyapunov function, it does not
necessarily play a kinetic role analogous to the one the energy plays in
equilibrium systems, and consequently the kinetics can usually not be
determined
from thermodynamics quantities of the bath.  In addition when more than one
gross
variable is considered and  when  the bath is not in thermal equilibrium the
reduced description {\it need not posses a local ``energy-type" function of the
gross variables in the   Langevin equations} (i.e. the  mean ``force" is not
necessarily curl free).  This is true in our example {\it even when the state
of
the of the bath can be described by a scalar thermodynamic quantity}, such as
in
the quasi-thermal situation discussed above.

I am particularly indebted to  Mark Dykman for many fruitful conversations and
suggestions.

\begin{figure}[t]
\vspace{1.45in}
\caption{\sf Typical dressed ratchet potential $\tilde U(Q)$.}
\end{figure}

\begin{figure}[t]
\vspace{1.45in}
\caption{\sf Physical information $u_i(\omega)$ density near zero frequency.
  The generic
cases where the phonon bath acts as an information source and the engine runs,
forward and  where it acts as a sink and the engine runs backward are shown. }
\end{figure}

\end{document}